\documentclass[letterpaper,twocolumn,10pt]{article}

\usepackage{usenix-2020-09}
\usepackage{graphicx}
\usepackage{algorithm}
\usepackage{algorithmicx}
\usepackage{paralist}
\usepackage{enumitem}
\usepackage{algpseudocode}
\usepackage{ifthen}
\usepackage{url}
\usepackage{comment}
\usepackage{listings}
\usepackage{balance}
\usepackage{xspace}
\usepackage{pdfpages}
\usepackage[sort,compress,noadjust]{cite}
\usepackage{hyperref}
\usepackage{cleveref}
\usepackage{dirtytalk}
\usepackage[font=small]{caption}
\usepackage{subcaption}
\usepackage{multirow}
\usepackage{multicol}
\usepackage{tikz}
\usepackage{booktabs}

\newcommand{\ATTACK}{Prefix Siphoning\xspace}
\newcommand{\Attack}{Prefix siphoning\xspace}
\newcommand{\attack}{prefix siphoning\xspace}

\newcommand*\circled[1]{\tikz[baseline=(char.base)]{
            \node[shape=circle,draw,inner sep=.7pt] (char) {#1};}}

\newif\iffullpaper
\fullpapertrue
\iffullpaper
\newcommand{\ShortPaper}[1]{\ignorespaces}
\newcommand{\FullPaper}[1]{#1}
\else
\newcommand{\ShortPaper}[1]{#1}
\newcommand{\FullPaper}[1]{\ignorespaces}
\fi

\makeatletter
\newcommand{\crefnames}[3]{%
  \@for\next:=#1\do{%
    \expandafter\crefname\expandafter{\next}{#2}{#3}%
  }%
}
\makeatother
\crefnames{part,section,subsection}{\S}{\S\S}
\crefnames{para,paragraph,subparagraph}{\P}{\P\P}
\crefnames{listing}{Listing}{Listings}
\crefnames{sublisting}{Listing}{Listings}
\crefnames{figure}{Figure}{Figures}
\crefnames{table}{Table}{Tables}
\crefnames{algorithm}{Listing}{Listings}
\crefnames{appendix}{Appendix}{Appendices}
\crefnames{char}{}{}

\newlist{chars}{enumerate}{1}
\setlist[chars]{label={\textbf{C\arabic*}},leftmargin=*,ref={C\arabic*}}
\crefalias{charsi}{char}

\makeatletter
\renewcommand{\section}{\@startsection{section}{1}{0pt}%
{-2ex plus -1ex minus -.2ex}{1.5ex plus.2ex}%
{\reset@font\large\bf}}
\renewcommand{\subsection}{\@startsection{subsection}{1}{0pt}%
{-1.5ex plus -1ex minus -.2ex}{1ex plus.2ex}%
{\reset@font\large\bf}}
\renewcommand{\subsubsection}{\@startsection{subsubsection}{1}{0pt}%
{-1ex plus -1ex minus -.2ex}{1ex plus.2ex}%
{\reset@font\bf}}
\renewcommand{\paragraph}{%
  \@startsection{paragraph}{4}%
  {\z@}{1.25ex \@plus 1ex \@minus .2ex}{-1em}%
  {\normalfont\normalsize\bfseries}%
}
\makeatother

\begin{document}

\date{}

\title{\Large \bf \ATTACK: Exploiting LSM-Tree Range Filters For Information Disclosure\FullPaper{ (Full Version)}}

\author{
{\rm Adi Kaufman\thanks{Both authors contributed equally to this research.}}\\
Tel Aviv University
\and
{\rm Moshik Hershcovitch\footnotemark[1]}\\
Tel Aviv University \& IBM Research
\and
{\rm Adam Morrison}\\
Tel Aviv University
}

\maketitle

\begin{abstract}

Key-value stores typically leave access control to the systems for which they act as storage engines.
Unfortunately, attackers may circumvent such read access controls via \emph{timing attacks} on the key-value store,
which use differences in query response times to glean information about stored data.

To date, key-value store timing attacks have aimed to disclose stored values and have exploited external mechanisms that can be disabled for protection.
In this paper, we point out that \emph{key} disclosure is also a security threat---and demonstrate key disclosure timing attacks that exploit mechanisms of the key-value store itself.

We target LSM-tree based key-value stores utilizing \emph{range filters}, which have been recently proposed to optimize LSM-tree range queries.
We analyze the impact of the range filters SuRF and prefix Bloom filter on LSM-trees through a security lens, and show that they enable a key disclosure timing attack, which we call \emph{\attack}.
\Attack successfully leverages benign queries for non-present keys to identify prefixes of actual keys---and in some cases, full keys---in scenarios where brute force searching for keys (via exhaustive enumeration or random guesses) is infeasible.

\end{abstract}

\section{Introduction}

Key-value stores serve as the storage engines of many cloud and enterprise systems, from object caches~\cite{ScalingMemcache,FollowFeed,NetflixCaching} through stream processing~\cite{RealtimeFacebook,Flink,Samza} to database systems~\cite{mongodb,hyperdex,CockroachDB,MyRocks}.
Performance of these modern data intensive systems often depends on their key-value storage engine's performance~\cite{PebblesDB}.
Consequently, research on key-value stores overwhelmingly focuses on \emph{efficiency}: from I/O efficiency of writes~\cite{Dostoevsky,LSM-Bush},  point queries~\cite{dayan2017monkey,DayanOptimal}, and range queries~\cite{Rosetta,SuRF} to
memory efficiency~\cite{SILT,SkimpyStash}, energy efficiency~\cite{FAWN}, multi-core scalability~\cite{cLSM,C4}, and reducing I/O write amplification~\cite{PebblesDB}.

But systems also depend on their key-value storage engine for the \emph{security} of stored data.
This dependency is not obvious, because key-value stores typically provide only a dictionary abstraction without access control mechanisms~\cite{RocksDB,redis,Leveldb,SplinterDB}, leaving access control to the system.
Systems enforce access control by mediating user accesses to the key-value store, often based on access control lists (ACLs) stored as value metadata in the key-value store.
While this approach blocks users from directly making unauthorized queries, users may still be able to indirectly glean information about restricted data if the key-value store is vulnerable to \emph{timing attacks}~\cite{TimingAttack}.

A timing attack exploits differences in query response times to glean information about stored data.
A system using a key-value store that is vulnerable to timing attacks can itself become vulnerable to such attacks, because the system's query response time depends on the storage engine's response time, making differences in key-value query response times manifest as differences in the system's response times.

To date, key-value store timing attacks~\cite{KVDataAttack1,KVDataAttack2} have aimed to disclose stored values.
We point out, however, that \emph{key} disclosure is also a security threat.
In some systems, keys can \emph{explicitly} contain secret data.
For example, database systems that use key-value storage engines (e.g., CockroachDB, YugabyteDB, or MyRocks) encode rows (or subsets of rows) onto keys~\cite{KVSforDB,CockroachDBuseRocks,MyRocksUseRocks,YugabyteUseRocks}.
This makes key disclosure equivalent to database data disclosure.
Keys may also be \emph{implicitly} secret, with users expecting them to be hard to obtain.
For instance, in object storage systems, such as Amazon S3, identifying valid keys may create an insecure direct object reference vulnerability~\cite{IDOR}, which enables attackers to probe access to the objects associated with the disclosed keys.

Unfortunately, resilience to timing attacks is not a goal in existing key-value efficiency work---in fact, such resiliency can be at odds with improved performance.
In this paper, we demonstrate this trade-off:
we analyze key-value store performance mechanisms through a security lens and show that they enable a key disclosure timing attack.

We focus on write-optimized key-value stores based on log-structured merge (LSM) trees~\cite{LSMTree}, which are in widespread use~\cite{RocksDBAnalysis,conway_et_al:LIPIcs:2018:9043,dayan2017monkey,PebblesDB,RocksDB,LSM-SSD,Chucky,Dostoevsky,chang2008bigtable,cassandra,BLSM,cLSM}.
In these designs, data in secondary storage consists of multiple immutable files called \emph{SSTables}.
LSM-trees can efficiently sustain write-intensive workloads, but queries may require multiple I/Os to search the many SSTables~\cite{LSMTree,BLSM}.
LSM-trees minimize unnecessary I/Os by issuing the I/O only if the queried key is likely to be in the SSTable.
Likelihood is determined by querying an in-memory \emph{filter}~\cite{BloomFilter}, which space-efficiently \emph{approximately} represents the SSTable's contents.
Specifically, filter queries can make ``one-sided'' errors: if the queried key is present in the SSTable, then the filter always returns true; but for a small fraction of non-present keys, the filter might return a false positive response.

Standard filters can answer point (single-key) queries~\cite{BloomFilter,CF,QF}, but do not support range queries of the form ``does the SSTable contain a key in range $\left[X, Y\right]$.''
Consequently, LSM-tree range queries must search the many SSTables, performing multiple superfluous I/Os~\cite{SuRF}.
To address this problem, recent work has proposed \emph{range filters}, which are filters that support range queries in addition to point queries.
Range filters such as SuRF~\cite{SuRF} and RocksDB's prefix Bloom filter~(PBF)~\cite{PBF} compactly store some or all prefixes of each of the SSTable's keys, and leverage this information to answer range and point queries.

From a security perspective, however, we show that \emph{certain range filters enable a key disclosure timing attack on LSM-trees}.
We describe an attack framework, called \emph{\attack}, which exploits general range filter characteristics present in both SuRF and PBF.
\Attack successfully leverages benign point queries for non-present keys to identify prefixes of actual keys---and in some cases, full keys---in scenarios where brute force searching for keys (via exhaustive enumeration or random guesses) is infeasible.

\Attack targets systems with the common design paradigm of storing a key's ACLs as part of its value~\cite{ceph,AmazonS3}, which means that to check access permissions, the system's query handling always tries to read the queried key's value from the key-value store.
\Attack exploits this property to determine if a random key is one on which the LSM-tree's filter returns a false positive.
This is possible because whether the filter returns true or false can be determined by the attacker observing the query's response time, as the filter's response decides whether the LSM-tree performs I/Os.
For range filters meeting our characterization, finding a false-positive key implies that the false-positive key shares a prefix with some stored key.
\Attack then performs further point queries---tweaking the queried key---to maximize the length of the disclosed prefix.
\Attack can sometimes subsequently perform a limited enumeration search to fully identify the stored key.
Our \attack implementation performs multiple such steps concurrently, ultimately extracting multiple keys or prefixes.

We evaluate \attack against SuRF and PBF analytically as well as empirically and demonstrate its feasibility in practice.
For example, we successfully use \attack to extract 64-bit stored keys from a RocksDB~\cite{RocksDB} datastore employing SuRF in minutes, whereas brute force search of this key space is infeasible.
Our analysis and evaluation also quantify the cost of \attack, showing that it effectively reduces the key search space size by multiple orders of magnitude.
For instance, SuRF \attack requires $\approx$\,10\,M queries to disclose a key from a 50\,M 64-bit key dataset---implying a $40992\times$ reduction of the key search space size.

Our results draw attention to the security vs. performance trade-offs in key-value store design, and encourage practitioners and researchers to evaluate the security impact of their work.
We hope that our characterization of vulnerable range filters will spur research on more secure filters.

\FullPaper{
\paragraph{Contributions}
We make the following contributions:
\begin{itemize}[leftmargin=*,itemsep=0pt]
\item \textbf{Problem (\cref{sec:intro_sc})}. We point out that keys can contain explicitly or implicitly sensitive data, creating the problem of key disclosure in key-value stores.
\item \textbf{\Attack (\cref{sec:side_attack,sec:impl}).} We define a general key disclosure attack template and characterize the range filter properties sufficient for mounting the attack.
\item \textbf{Attacking SuRF and PBF (\crefrange{sec:attack-surf}{sec:attack-pbf}).} We describe\FullPaper{ and analyze} instantiations of \attack against SuRF and PBF.
\item \textbf{Evaluation (\cref{sec:evaluation_sc}).} We empirically show the feasibility of \attack against a full-blown RocksDB store employing SuRF and PBF.
\end{itemize}
}

\section{Background}\label{sec:side_contribution}

This section provides background on key-values stores~(\cref{key-value store}), LSM-trees~(\cref{LSM store}), and filters~(\cref{filters_background}).

\subsection{Key-value stores} \label{key-value store}

A key-value store exposes a dictionary-like abstraction with the following operations.

\begin{itemize}[leftmargin=*,itemsep=0pt]
\item
\emph{put}($k,v$). A \emph{put} stores a mapping from key $k$ to value $v$.
If key is already present in the store, its value is updated.

\item
\emph{get}($k$). The \emph{get()} (or point query) returns the value associated with the requested key.

\item
\emph{range\_query}($k_1,k_2$). A range query returns all key-value pairs falling within the given range.
\end{itemize}

Due to their simple and general abstraction as well as high performance, key-value stores serve as the storage engines for many, more complex systems.
Examples of such systems include database systems (e.g., Cassandra~\cite{cassandra}, MongoDB~\cite{mongodb}, and MySQL~\cite{mysql}) and storage systems (e.g., CEPH~\cite{ceph}).

\subsection{LSM-based data stores} \label{LSM store}

The log-structured merge (LSM) tree~\cite{LSMTree} is a popular choice as the core storage structure for write-optimized key-value stores, which must sustain write-intensive workloads.
An LSM-tree consists of levels, each of which contains multiple immutable static sorted table (SSTable) files storing key/value pairs.
Two SSTs at the same level never overlap in the key range they store, but SSTables at different levels may overlap.

A \emph{put} request inserts the key-value pair into an in-memory buffer called the Memtable, which is the LSM-tree's only mutable storage object.
Once the Memtable fills up, its data is flushed to secondary storage as an SSTable file.
The LSM-tree periodically performs \emph{compaction}, where it unifies SSTs between levels to eliminate duplicate (stale) key-value pairs.

A \emph{get} query searches for the target key in a top-down manner: first in the Memtable and subsequently in the relevant SSTable (if it exists) in each level.
Searching an SSTable requires I/Os to read it from secondary storage.
Once the key is found, its value is returned and the query completes.

However, this design penalizes queries, which may require multiple I/Os to search many SSTables~\cite{LSMTree,BLSM}.
In particular, a \emph{get()} for a \emph{non-present} key (not associated with any value) searches every level before failing.
This not only increases the query response time, but may ``thrash'' the page cache by reading in many SSTables which will not be accessed later.

LSM-trees minimize unnecessary I/Os by issuing the I/O only if the queried key is likely to be in the SSTable.
Likelihood is determined by querying an in-memory filter (described in~\cref{filters_background}), which space-efficiently \emph{approximately} represents the SSTable's contents.
\FullPaper{Specifically, filter queries can make ``one-sided'' errors: if the queried key is present in the SSTable, then the filter always returns true; but for a small fraction of non-present keys, the filter might return a false positive response.}
The LSM-tree only reads an SSTable from secondary storage if its filter returns true for the queried key.
As a result, most non-present key queries can respond without performing I/Os.

Likewise, a range filter~(\cref{back:range-filters}) can answer both point and range queries with one-sided errors.
Using a range filter instead of a standard filter enables an LSM-tree to avoid superfluous I/Os also for range queries, which can improve range query throughput by orders of magnitude~\cite{Rosetta}.

\subsection{Filters}\label{filters_background}
A \emph{filter}~\cite{BloomFilter} is a data structure used to approximately represent a set $D$ a of keys.
A filter can be \emph{immutable} or \emph{dynamic}.
An immutable filter is provided $D$ upon its creation and can subsequently only be queried.
A dynamic filter learns $D$ dynamically, via \emph{insert} operations.

Responses for filter queries allow ``one-sided'' errors: if $k \in D$, then a query for $k$ returns true;
but for a fraction of keys $k \not\in D$, a query for $k$ might answer true instead of false.
We say $k$ is a \emph{positive}/\emph{negative key} if a filter a query for $k$ answers true or false, respectively.
A positive key $k$ is a \emph{false positive} if $k \not\in D$.
We also say that the filter \emph{passes} positive keys and \emph{rejects} negative keys.

Filters are compared by their space efficiency and false-positive rates.
Space efficiency is measured in bits per key.
The \emph{false-positive rate} (FPR) of a filter is the probability over keys not in $D$ of being a false positive.
I.e., $FPR = FP/(FP+NK)$, where $FP$ is the number of false-positive keys and $NK$ is the number of negative keys.
Filters typically have configurable FPRs, with lower FPRs requiring more bits per key for increased accuracy~\cite{BloomFilter,CF,QF}.

\FullPaper{
Because filters are approximate, they are more space- and time-efficient than exact set representations (indexes).
These properties enable maintaining in-memory filters for large datasets, such as an LSM-tree's key set.
}

\paragraph{Bloom filters}
A Bloom filter~\cite{BloomFilter} is a widely-used dynamic filter (e.g., the default filter of RocksDB).
It consists of an $m$-bit array and $j$ hash functions $H_1,\dots,H_j$. The parameters $m$ and $j$ determine the filter's FPR and space.
Insertion of key $k$ sets the bit indexes $H_1(k),\dots,H_j(k)$.
A query for $k$ returns true if and only if all bit indexes $H_1(k),\dots,H_j(k)$ are set.

\subsubsection{Range filters} \label{back:range-filters}

A \emph{range filter} is a filter that also supports range queries with one-sided error:
a query for $[a,b]$ returns true if there exists $k \in D \cap [a,b]$, but might also return true if $D \cap [a,b]$ is empty.

\FullPaper{
LSM-trees use range filters to quickly identify when there are no keys from a range in the dataset $D$,
which allows an LSM-tree \emph{range\_query} to respond without performing I/Os.
}

\section{Motivation: avoiding key disclosure} \label{sec:intro_sc}

We observe that keys stored in a key-value storage engine can contain sensitive data.
It is therefore desirable that users are not able to efficiently discover stored keys that they are not authorized to access.
Of course, users can always guess such keys and check if their queries return an authorization error, but such brute force searches are infeasible on large
key spaces.
The goal is for brute force search to be the only attack option, i.e., to block more efficient key extraction attacks.

\paragraph{Explicitly secret keys}
Some systems encode secret data in stored keys, which makes key disclosure equivalent to disclosure of the encoded data.
For example, database systems such as CockroachDB, YugabyteDB, and MyRocks store table rows as values in a key-value storage engine,
with the associated key consisting of the table's id and the row's primary key (one of the cell values).
The motivation for this technique is that it enables the database system to perform efficient primary key lookups using key-value store range queries~\cite{KVSforDB,CockroachDBuseRocks,MyRocksUseRocks,YugabyteUseRocks}.

\paragraph{Implicitly secret keys}
In many cases, keys are tacitly assumed to be secret or, at least, hard to guess.
One example of implicitly secret keys are \emph{object identifiers}.
Many web applications and object storage systems maintain object id-to-value mappings in a key-value store.
Key disclosure thus allows attackers to probe access to the associated objects, resulting in an insecure direct object reference vulnerability~\cite{IDOR}.
While objects typically have ACLs, users often neglect to configure these ACLs.
This is not a hypothetical concern: for instance, there are numerous scanning tools for ``open'' (unprotected) Amazon S3 objects~\cite{S3Scanner1,S3Scanner2,S3Scanner3,S3Scanner4,S3Scanner5,S3Scanner6},
and open S3 objects have led to exfiltration of employee information, personal identification information, and other sensitive data~\cite{S3breach}.

\section{Threat model} \label{sec:threat}

We consider a \emph{high-level system}, such as a database system or object store, that utilizes a key-value storage engine to respond to user queries.
Key ACLs are stored as part of the value associated with the key.
As the high-level system performs key-value queries to satisfy a user's query, it checks the ACL of each key it accesses by inspecting the key's value.
If the user is not authorized to read a key, the system returns a failure response to the user.

The attacker's goal is to identify keys stored in the system's key-value storage engine.
The attacker cannot compromise the system (e.g., to run attack code) and cannot eavesdrop on requests performed by other users and/or on their responses.
The attacker can only interact with the system by making requests via its interfaces, such as a representational state transfer (REST) API~\cite[Chapter 5]{fielding-phd}.

We assume that the attacker can craft their requests in a way that causes the high-level system to make key-value store point queries for arbitrary keys (i.e., chosen by the attacker) while processing the request.
For simplicity, we refer to this process as  the attacker ``querying the key-value store.''

We make no assumption about the attacker's physical location with respect to the attacked system.
We only assume that the attacker can observe microsecond-level timing differences in the response times of queries for different keys.
Prior work has shown that this assumption is true for attacks over both local and wide area networks.
For instance, Crosby et al. were able to measure a difference of 20\,$\mu$s over the circa 2009 Internet (and 100\,ns over a local area network)~\cite{TimingAttackData}.
This ability can be improved in specific cases.
When attacking a system hosted in the public cloud, for example, the attacker can turn themselves into a local-area attacker by
placing themselves in the datacenter hosting the target.
Moreover, systems that process different requests concurrently (e.g., HTTP/2 servers) are vulnerable to concurrency-based timing attacks~\cite{TimingAttack2}, which can observe timing differences of 100\,ns over the Internet.

\section{\Attack} \label{sec:side_attack}

\Attack is a general template for conducting timing attacks, extracting partial or full keys, on systems that use an LSM-tree based storage engine with a certain type of \emph{vulnerable} range filter
(for both point and range queries).
The class of vulnerable range filters contains the filters SuRF~\cite{SuRF} and RocksDB's prefix Bloom filter~(PBF)~\cite{PBF}.

\Attack exploits range filters that respond to point queries based on key prefix information, which exists to support range queries---i.e., filters where range query support affects the point query implementation.
Accordingly, \attack is based \emph{only on point queries} and does not perform range queries.
Henceforth, therefore, the term ``query'' always refers to a \emph{get()} point query.
We leave exploring attacks against range queries to future work.

In the following, we describe the attack's high-level ideas~(\cref{sec:side_attack:idea}), characterize the class of vulnerable filters~(\cref{subsubsec:vulnrableFilters}), and present the attack template~(\cref{sec:side_attack:template}).
We describe instantiations of the attack against SuRF and PBF in~\crefrange{sec:attack-surf}{sec:attack-pbf}.

\paragraph{Notation}
We treat keys as sequences of symbols over an alphabet $\Sigma$ (e.g., bytes).
When $x$ denotes a key or a set, then $|x|$ refers to the number of symbols or elements, respectively, that $x$ contains.

\subsection{High-level ideas} \label{sec:side_attack:idea}

\Attack exploits an \emph{inherent} trait of filter use in LSM-trees: that whether a key ``passes'' the filter determines if the LSM-tree searches the SSTable for the key to satisfy a query.
This means that for SSTable files that do not reside in the OS page cache, the filter's output for a key significantly affects the LSM-tree's query response time.
If the filter returns false for the key, the response is satisfied with only main memory access; otherwise, the LSM-tree needs to perform I/Os to read SSTables from secondary storage.
Even for fast storage such as NVMe devices, the difference in query response times between these two cases is enough to affect the system's overall response time in an attacker-measurable way.

This basic filter trait suffices to mount an ``approximate membership test'' timing attack.
The attack simply queries for the target key $k$ and measures the response time.
If the response time is fast (i.e., $k$ is rejected by the filters), then $k$ is definitely not stored in the LSM-tree.
Otherwise (i.e., $k$ passes some filter), then $k$ is likely in the LSM-tree.
The key $k$ might also be a filter false positive and not exist in the LSM-tree, which occurs with a probability bounded by the filter's FPR.

\Attack starts by randomly generating keys until it finds a key that ``passes'' the membership test above.
For random keys, passing the test overwhelmingly means that the key is a filter false positive.
Crucially, it takes just hundreds of attempts to find a false-positive key, because filters are typically configured for FPRs of a few percents for space efficiency reasons~\cite{SuRF}.

Our main observation is that in vulnerable range filters, a false-positive key likely shares a prefix with some stored key $k$, whereas negative keys (rejected by the filter) do not (at least with high probability).
The crux of a \attack attack is an algorithm exploiting this trait to \emph{identify} the shared prefix $k'$ through $O(|k|)$ further queries for modified keys iteratively derived from the initial false-positive key.

The revealed prefix of $k$ can already contain sensitive information.
But if the system's query responses distinguish between failures due to target key non-presence and lack of authorization, \attack can fully extract $k$ by performing brute force search of the unknown suffix, thereby \emph{extending} the revealed prefix to $k$.

Of course, a system whose responses distinguish between non-present and unauthorized keys is also vulnerable to ``brute force'' key guessing or enumeration attacks based using the above ``membership test'' primitive.
But such attacks are infeasible for many key spaces (e.g., 64-bit or string keys).
The point of \attack is to narrow down the search space by exploiting vulnerable range filters.
Moreover, \attack extracts key prefixes even if the target system's responses do not reveal whether a key is non-present or unauthorized, whereas the ``membership test'' primitive cannot.

\subsection{Vulnerable range filter characterization} \label{subsubsec:vulnrableFilters}

We denote an instance of the filter by $F$ and the set of keys it represents by $D$.
A range filter is \emph{vulnerable to \attack} if it has the following characteristics, denoted~\crefrange{char:1}{char:2}.
They say that a false-positive key $\kappa$ likely shares a prefix with some key from $D$ and that an attacker can efficiently identify this prefix by making queries for keys derived from $\kappa$.

\begin{chars}
\item \label{char:1} If $\kappa$ is a false-positive key for $F$, then with high probability, $\kappa$ shares a prefix with some $k \in D$.
\item \label{char:2} There exist the following probabilistic algorithms, which work by querying the system:
\begin{enumerate}[leftmargin=*]
\item \emph{FindFPK()}: Using an expected constant number of queries, outputs a random false-positive key $\kappa$.
\item \emph{IdPrefix($\kappa$):} Given a false-positive $\kappa$, uses $O(|\kappa|)$ queries to identify the shared prefix $k'$ that $\kappa$ shares with some key $k \in D$, if such a prefix exists; otherwise, the output is unspecified.
\end{enumerate}
\end{chars}

The \emph{FindFPK} and \emph{IdPrefix} algorithms are specific to the range filter design, and need to be developed by the attacker.%
\footnote{Existence of \emph{FindFPK} and \emph{IdPrefix} is required in addition to~\cref{char:1} because a filter satisfying only~\cref{char:1} may not allow an attacker to extract the prefixes.}
We refer to designing such algorithms for a range filter as \emph{instantiating} the attack against that filter.

\Cref{char:2} implies existence of a timing attack, and is therefore formally sufficient to characterize the vulnerability.
In practice, however, our attack instantiations rely on fundamental properties of filter use in LSM-trees.
To highlight this aspect of the attacks, we explicitly capture these properties in~\cref{char:distinguish}.

\begin{chars}[resume]
\item \label{char:distinguish}\label{char:fpk}
\begin{enumerate}[leftmargin=*]
\item A \emph{get}($k$) query's response time is measurably lower if $k$ misses in every filter than if $k$ hits in some filter.
\item The filter's FPR is small but non-negligible (e.g., 1\% or 0.1\%).
\end{enumerate}
\end{chars}

\cref{char:distinguish}(1) implies that it is possible to distinguish negative from positive keys using query response times.
It is trivially true because LSM-trees employ filters to speed up queries for which SSTable searching is superfluous, such as filter misses.
Our attacks in this paper exploit microsecond-level time differences between queries satisfied completely from main memory and those that require I/O to secondary storage.
(There remain time differences between queries that read an in-memory SSTable residing in the OS page cache and those that do not, due to a filter miss.
We leave exploiting such smaller time differences to future work.)

\cref{char:distinguish}(2) implies that generating keys uniformly at random will generate a false-positive key with hundreds to thousands of attempts, on average.
It holds because in practice, filters are typically configured with small but non-negligible FPRs (e.g., $0.5\%$--$5$\%), as negligibly small FPRs blow up the filter's memory consumption~\cite{SuRF}.%
\footnote{\Attack can still be performed for exponentially low false positive rates, but its cost (in terms of number of queries needed) increases proportionally to the decrease in the false positive rate.}

\subsection{\Attack template} \label{sec:side_attack:template}

\Attack consists of two phases. First, a preliminary phase learns to distinguish queries of negative and positive keys~(\cref{sec:attack:step1}).
The second phase consists of multiple rounds, each of which extracts a key or key prefix~(\cref{sec:attack:step2}).
Rounds are run concurrently (see~\cref{sec:impl}).

\subsubsection{Learning to distinguish positive from negative keys} \label{sec:attack:step1}

The attack starts with a preliminary phase that builds a distribution of query response times, which is used by the second phase to distinguish positive from negative keys.

The distribution is built by measuring response times of multiple \emph{get()} requests for random keys.
With large key spaces, such random keys are mostly negative keys, but a small (though non-negligible) fraction will be positive (due to~\cref{char:fpk}).
Such positive keys are overwhelmingly likely to be false positives, but that does not matter for this step, which is only concerned with distinguishing negative from positive keys, regardless of whether the positive output is correct.

The expected distribution observed is a bimodal distribution, with peaks corresponding to the average response time of negative and positive keys.
From this distribution, the attacker can derive a cutoff value that likely distinguishes negative (fast) from a positive (slow) queries.

\subsubsection{Extracting keys} \label{sec:attack:step2}

This phase consists of multiple rounds, each of which extracts a key.
Each round consists of three steps: \circled{1} finding a false-positive key $\kappa$, \circled{2} identifying the prefix that $\kappa$ shares with some stored key $k$,
and, when possible, \circled{3} extending the prefix to extract $k$.
Rounds are run concurrently~(\cref{sec:impl}).

Step~\circled{1} and \circled{2} simply invoke the attacker's \emph{FindFPK} and \emph{IdPrefix} algorithms, respectively.
These steps are actually the ``meat'' of the attack, and we later describe their instantiations for SuRF~(\cref{sec:attack-surf}) and RocksDB's prefix Bloom filter~(\cref{sec:attack-pbf}).

Whether step~\circled{3} is possible depends on the properties of the attacked system (and this is why it is not part of the vulnerable range filter characterization).
If the system's query responses distinguish between failures due to target key absence and lack of authorization, then the attacker can extend the revealed prefix $k'$ with some symbol sequence $\alpha$
and query for the key $k' \, \alpha$. The response will indicate lack of authorization if and only if $k' \, \alpha$ is a valid key.
The attacker can thus iterate over all possible suffixes until $k$ is found.
Because $k$ is not known to the attacker, they must first try all possible single symbol extensions, then all two symbol extensions, and so on.
This process requires $O(|\Sigma|^{|k|-|k'|})$ queries, which can be several orders of magnitude less than a full-key brute force search.
Crucially, step~\circled{3} only attempts to extend ``long'' prefixes, for which extension is feasible. Other prefixes are discarded.

\paragraph{Rationale for step decomposition}
For fixed-length keys, it might seem that the \emph{IdPrefix} algorithm (step~\circled{2}) for identifying the prefix is superfluous.
After all, given that $\kappa$ shares a prefix with some stored key $k$, the attacker can enumerate all possible suffixes from the end to the beginning, until identifying $k$.
For example, suppose keys are 14-character strings and the attacker has found a false-positive key \texttt{manchestercars} because it shares the prefix \texttt{manchesterc} with the stored key \texttt{manchestercity}.
Without knowing (or caring about) the shared prefix, the attacker can start querying for \texttt{manchestercara}, \texttt{manchestercarb}, $\dots$, \texttt{manchestercaaa}, \texttt{manchestercaab}, and so on---all of which fail due to key absence---until reaching \texttt{manchestercity}, which will fail due to lack of authorization.
As before, this process requires $O(|\Sigma|^{|k|-|k'|})$ queries and so it theoretically achieves the same results directly, without requiring an \emph{IdPrefix} algorithm.

Why, then, is existence of an \emph{IdPrefix} algorithm defined as one of the characteristics of a vulnerable filter?
The answer is that without knowledge of the prefix, the attacker cannot efficiently schedule their work in step~\circled{3}.
They cannot distinguish a small suffix space (as in the example above) from a huge space---e.g., if the false-positive key only shared the prefix \texttt{m} with \texttt{manchestercity}.

The \emph{IdPrefix} algorithm protects us from the above pitfall.
By identifying the shared prefix, it enables the attacker to decide whether to try and extend the prefix to a full key.
Moreover, when multiple rounds execute concurrently, the attacker can collect many prefixes and then prioritize extending the longest ones.

\section{SuRF \attack} \label{sec:attack-surf}

Here, we instantiate a \attack attack against LSM-trees employing the SuRF~\cite{SuRF} range filter.
\Cref{sec:surf-primer} summarizes SuRF and~\cref{sec:surf-vuln} shows that it is vulnerable to \attack.

\subsection{SuRF primer} \label{sec:surf-primer}

The succinct range filter (SuRF)~\cite{SuRF} is the first proposed general range filter.
Like the LSM-tree SSTables it approximates, SuRF is an immutable structure.
SuRF can speed up LSM-tree range queries by $5\times$, but it imposes a modest cost on point queries due to having higher FPRs than a Bloom filter~\cite{SuRF}.

\begin{figure}[t]
\vspace{-8mm}
\centerline{\includegraphics[width=\columnwidth]{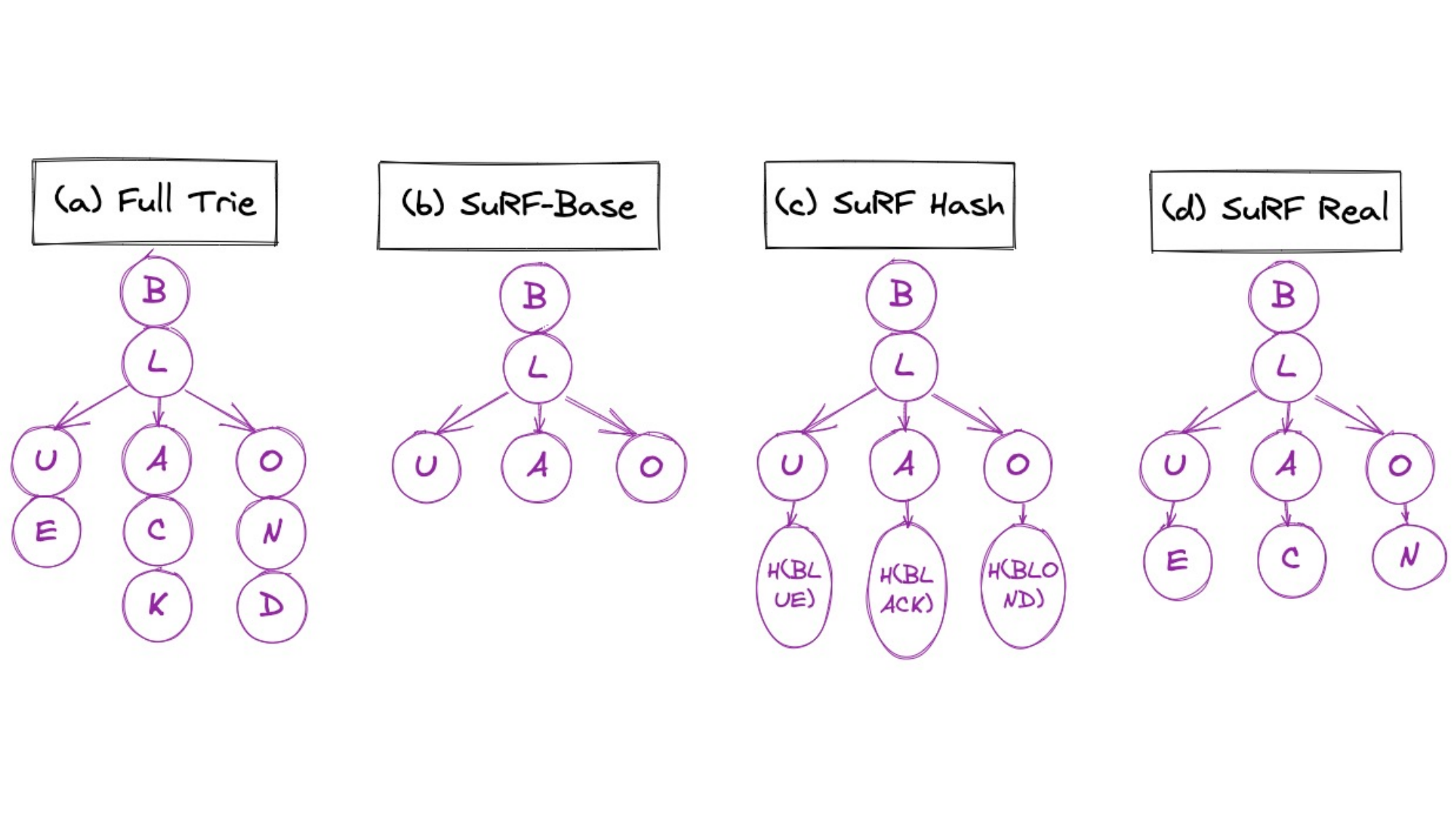}}
\vspace{-9mm}
    \caption{Trie and SuRF variants over the key set \texttt{BLUE}, \texttt{BLACK}, and \texttt{BLOND}. (Figure adapted from~\cite{SuRF}.)}
    \label{fig:surf}
\end{figure}

At a high level, SuRF is a pruned trie. A \emph{trie} is a tree data structure that stores keys sorted according to the lexicographic order of $\Sigma$.
Each edge is labeled with a symbol and each node corresponds to the concatenation of all edge labels on the path to that node.
Each leaf thus corresponds to a key and each internal node to a key prefix (\Cref{fig:surf}(a)).
An internal node can also correspond to a key (if the key set is not prefix-free), which is indicated by one of its fields.
For space-efficiency, SuRF uses a succinct trie representation.

SuRF further saves space by \emph{pruning} the trie. The basic SuRF variant (SuRF-Base) stores the minimum length key prefixes that uniquely identify each key,
i.e., shared key prefixes plus the symbol following the shared prefix of each key (Figure~\ref{fig:surf}(b)).
SuRF's pruning results in a space-efficient but only approximate representation of the key set.

Both point and range queries are satisfied from the pruned trie structure.
A \emph{get}($k$) returns true (possibly erroneously) if and only if the path induced by $k$ terminates at a node associated with a key.
For example, in Figure~\ref{fig:surf}(b), \texttt{BLOOD} is a false positive.
Range queries rely on the trie's ordered structure.
For example, to check if the SuRF contains a key $k \in [a,b]$, the query finds the node corresponding to the smallest key $\geq a$.
If it corresponds to a key $> b$, the query returns false; otherwise, it returns (possibly erroneously) true.

\paragraph{SuRF variants to reduce FPR}
SuRF-Base's FPR is data-dependent, i.e., depends on the key set. Compare, for example, two sets of 26 keys: $A = \left\{ x \, \alpha \, | \, x \in A,\dots,Z \right\}$ and $B = \left\{ \alpha \, x \, | \, x \in A,\dots,Z \right\}$, where $\alpha$ is some long string.
For $A$, SuRF's FPR is nearly 100\%, as any key except $A,\dots,Z$ is a false positive. But for $B$, the FPR is extremely small, as only keys that begin with $\alpha$ pass the filter.

To improve the FPR, SuRF offers variants that augment SuRF-Base's pruned structure with a few bits per leaf of information about the leaf's suffix.
These bits reduce the FPR by allowing queries to reject keys that share a prefix with the stored key but have a different suffix, in exchange for increasing per-key memory consumption.

\emph{SuRF-Hash}~(\cref{fig:surf}(c)) hashes the leaf's key and stores $n$ bits from the hash value, where $n$ is configurable.
\emph{SuRF-Real}~(\cref{fig:surf}(d)) stores the first $m$ bits of the key's suffix, where $m$ is configurable.

\subsection{Vulnerability of SuRF} \label{sec:surf-vuln}

Every SuRF variant has the characteristics defined in~\cref{subsubsec:vulnrableFilters}.
\Cref{char:distinguish}(1) holds trivially.
\Cref{char:fpk}(2) holds empirically: SuRF-Base has an FPR of 4\% for random 64-bit keys and SuRF-Hash reduce this FPRs to $\approx 0.1$\%~\cite{SuRF}.
\Cref{char:1} holds because in every SuRF variant, \emph{every} false-positive key $\kappa$ shares a prefix with some stored $k$---\cref{char:1} holds with probability 1.

To show that~\cref{char:2} holds, we describe how to efficiently find a false-positive key~(\cref{sec:surf-fpk}) and how to identify the prefix that it shares with a stored key~(\cref{sec:surf-extend}).
We assume the ability to check if a key is a filter positive or negative key based on measuring query response times.
The implementation of this check is described in~\cref{sec:impl}.

\subsubsection{Finding a false-positive key (FindFPK)} \label{sec:surf-fpk}

For SuRF, our \emph{FindFPK} algorithm simply generates queries for uniformly random keys until it detects a positive response, based on the cutoff determined in the attack's preliminary learning phase~(\cref{sec:attack:step1}).
Due to~\cref{char:fpk}, this step is expected to terminate with a few hundreds to thousands of attempts.

We refer to the random positive key found as a false-positive key, because that is the overwhelmingly likely event.
However, the attack still works if, unbeknownst to the attacker, the found key is actually a true positive key.

\subsubsection{Identifying a shared prefix (IdPrefix)} \label{sec:surf-extend}

For a false-positive key $\kappa$, let $k = k(\kappa)$ be the stored key whose shared prefix $k'$ with $\kappa$ is the longest among all stored keys.
We write $\kappa = k' \alpha$ and $k = k' \beta$.
Our algorithm will output $k'$.

\paragraph{SuRF-Base/Real}
To find $k'$, we exploit SuRF's structure, namely that any key starting with a proper prefix of $k'$ is a negative key.
Let $\kappa = \kappa_1 \dots \kappa_n$.
We repeatedly remove the last symbol from the key, iteratively checking if the keys $\kappa_1 \dots \kappa_{n-1}, \kappa_1 \dots \kappa_{n-2}, \dots$ are negative or positive keys.
These keys will be positive until we remove a symbol from $k'$. Thus, the key checked before a negative key is found is $k'$.

If the attacked system does not support variable-length keys, removing symbols is not possible.
In this case, instead of removing symbols, we change them.
We iteratively check if the keys $\kappa_1 \dots \kappa_{n}', \kappa_1 \dots \kappa_{n-1}' \kappa_n\, \dots$ are negative or positive keys, where $\kappa_i' \neq \kappa_i$.
Similarly to before, if the first negative key found is $\kappa_1 \dots \kappa_j' \dots \kappa_{n}'$ then $k' = \kappa_1 \dots \kappa_j$.

Overall, the number of requests made is $O(|\kappa|)$.

\paragraph{SuRF-Hash}
SuRF-Hash complicates the attack, because modifying $\kappa$'s suffix can change its hash value, leading to a key that is rejected by SuRF despite sharing the prefix $k'$.
To address this problem, we assume SuRF's hash function $hash$ is public knowledge.
(This is a reasonable assumption, because the hash function's purpose is to reduce the FPR and not for security.)
We perform essentially the same algorithm(s) as for SuRF-Base/Real, but we only query each modified key $\kappa'$ if $hash(\kappa')=hash(\kappa)$.
We are still essentially assured to find keys to query, because SuRF-Hash stores only a small subset of the hash bits, for space-efficiency reasons.
For example, with the recommended 4 hash bits~\cite{SuRF} and using 8-bit symbols, on average 1 in 16 symbols tried will yield a hash collision and thus a key usable by the \emph{IdPrefix} algorithm.

Similarly, when trying to extend an identified prefix to a full key (step~\circled{3} in~\cref{sec:attack:step2}), we can skip querying any candidate key whose hash does not match the false-positive key's hash.

\FullPaper{
\subsection{Complexity analysis}\label{sec:stats_sc}

We analyze the case of uniformly random keys, which is the worst-case for our attack.
(If the key distribution is skewed, then~(1)~the guessing and full-key extraction steps can incorporate this knowledge; and~(2)~the prefixes SuRF stores are longer, so our attack will identify longer prefixes and thus extends them to full keys faster.)

For now, we assume $\Sigma=\{0,1\}$.
Consider a dataset $D$ of $n$ uniformly random $m$-bit keys.
The key factor determining \attack's effectiveness is the probability of \emph{FindFPK} guessing an \emph{exploitable} key $k$, which is~(1)~a SuRF false positive and~(2)~whose longest shared prefix with keys in $D$ is of length $l$, where $l$ is a predetermined constant for which it is feasible to extend $k$ into a full key~(\cref{sec:attack:step2}, step~\circled{3}).
Let $p^*=p(l,n,m)$ be the probability of this event.

To compute $p^*$, we first define the following
events, conditioned on $x \in D$ for an $m$-bit key $x$. Let $NoSP^{\geq l}_x$ be the event that $x$ has no shared prefix of length $\geq l$ with any other key in $D$.
Then $P(NoSP^{\geq l}_x) \approx (1-\frac{1}{2^l})^{n-1}$.%
\footnote{This is an approximation because it considers sampling of $n$ keys with repetitions. In the realistic case of $n \ll 2^l$, however, the probability of a repetition is negligible, so this approximation is essentially an equality.}
Let $SP^{\geq l}_x$ be the event that $x$ has a shared prefix of length $\geq l$ with some key in $D$.
Then $P(SP^{\geq l}_x)=1-P(NoSP^{\geq l}_x)$.
Now, let $SP^{=l}_x$ be the event that $x$'s longest shared prefix with other keys in $D$ is of length $l$.
Then $SP^{=l}_x = SP^{\geq l}_x \cap NoSP^{\geq l+1}_x$.
By the inclusion-exclusion principle,
$P(SP^{\geq l}_x \cap NoSP^{\geq l+1}_x) = P(SP^{\geq l}_x) + P(NoSP^{\geq l+1}_x) - P(SP^{\geq l}_x \cup NoSP^{\geq l+1}_x)$.

Denote the length of $x$'s longest shared prefix with other keys in $D$ by $maxPfx(x)$.
Observe that $P(SP^{\geq l}_x \cup NoSP^{\geq l+1}_x)=P(maxPfx(x)\geq l\textrm{ or }maxPfx(x)\leq l)=1$.
It follows that $P(SP^{=l}_x)=(1-\frac{1}{2^{l+1}})^{n-1} - (1-\frac{1}{2^{l}})^{n-1}$.

Using the law of total probability, we have that
$p^* = \sum_{x \in \{0,1\}^m} P(x \in D) \cdot P(SP^{=l}_x) \cdot \frac{1}{2^l} \cdot P(\textrm{SuRF}_x)$, where $P(\textrm{SuRF}_x)$ is that probability that, conditioned on $SP^{=l}_x$ happening, a uniformly random $k$ matching $x$ on its first $l$ bits is also a SuRF false positive.
(For example, for SuRF-Base with $\Sigma=\{0,1\}$, $P(\textrm{SuRF}_x)=1/2$ because $k$ needs to match $x$ on bit $l+1$.)
Finally, we obtain
$p^* = \frac{n}{2^l} \cdot P(SP^{=l}_x) \cdot P(\textrm{SuRF}_x)$,
because $P(x \in D)=n/2^m$.

Once an exploitable key $k$ is found by \emph{FindFPK}, its $l$-bit prefix is output by \emph{IdPrefix}. By~\cref{char:2}, \emph{IdPrefix} performs $O(m)$ queries.
In cases where the attacked system's query responses distinguish queries for absent and unauthorized keys, \attack proceeds to extract $k$ by enumerating all possibilities for $k$'s last $m-l$ bits, each time querying for the resulting key, until obtaining an ``unauthorized'' response, which implicitly discloses $k$. (Equivalently, the search can perform random guessing without repetition of guesses.)

The number of queries until $k$ is extracted in this manner can be modeled as a random variable with a negative hypergeometric distribution.
The negative hypergeometric distribution describes the number of elements randomly drawn from a population of size $N$, of which $K$ elements are considered ``successes'' and the rest are ``failures'', until $r$ failures are encountered.
The expected value of a negative hypergeometric random variable is $\frac{rK}{N-K+1}$.
\Attack's key extraction search can be modeled as such a process with $N=2^{m-l}$, $K=2^{m-l}-1$ (all possibilities except for $k$), and $r=1$.
Thus, the expected number of queries it makes is $(2^{m-l}-1)/2$.

Overall, the expected number of queries for \attack to extract a single key is $1/p^* + O(m) + (2^{m-l}-1)/2$.

\paragraph{Comparison to brute force searching}
Brute force enumeration (or guessing) of the entire key space in search of a key in $D$ can be modeled as the trial described by a negative hypergeometric distribution with parameters $N=2^m$, $K=2^m-n$, and $r=1$.
Therefore, the expected number of queries for a brute force search to identify a key in $D$ is $(2^m-n)/(n+1)$.

For perspective, we proceed to analyze and compare the attacks for 64-bit keys.

\paragraph{SuRF-Real}
\Cref{table:vulnerability_surf_real} shows the probability $p^*$ of a \emph{FindFPK} randomly guessed key being exploitable against SuRF-Real for 64-bit keys with varying $l$ and $n$ values. In addition, each cell
shows (in parentheses) the relative reduction in the number of queries
required to extract a key with \attack compared to a brute force search.
Table values are derived analogously to the above calculations, but accounting for SuRF working with bytes instead of bits in practice.

Observe that \attack becomes more effective as the dataset size grows, because the probability of guessing an exploitable key increases.
Moreover, we see that $l=40$ is a ``sweet spot'' for \attack.
It has a non-negligible $p^*$ and the subsequent key extension requires searching a manageable space of size $2^{24}$. Shorter $l$ values increase the key extension search space size, whereas for longer values, finding prefixes to extend becomes unlikely.

\paragraph{SuRF-Hash}
The analysis of SuRF-Hash is similar to that of SuRF-Real, except for accounting for differences in the factor $P(\textrm{SuRF}_x)$ in the computation of $p^*$.
For simplicity, assume SuRF-Hash uses 8-bit hash values. Then SuRF-Hash's results for $l$ are identical to SuRF-Real's results for $l+8$ (assuming that symbols are bytes). The reason is that for 8-bit hash values, the factor $P(\textrm{SuRF}_x)$ for both SuRF variants is $1/256$. In SuRF-Real, this is the probability of $x$ and $k$ matching on the $l+1$ byte. In SuRF-Hash, this is the probability that $x$ and $k$ have the same 8-bit hash value. In SuRF-Real, however, the prefix obtained if of length $l+8$, because if the additional byte stored in SuRF's trie. In contrast, SuRF-Hash stores the hash value and therefore the prefix obtained is of length $l$.

\begin{table}[t]
\begin{center}
\begin{tabular}{@{}lccccl@{}}\toprule
      \multicolumn{1}{c}{\textbf{prefix length ($l$)}} &
      \multicolumn{3}{c}{\textbf{dataset size ($n$)} }\\
       &    \multicolumn{1}{c}{\textbf{10\,M}} &
      \multicolumn{1}{c}{\textbf{50\,M}} &
      \multicolumn{1}{c}{\textbf{100\,M}} \\ \toprule
     \multicolumn{1}{c}{32 bits}  & 0.001 & 0.0006 & $6*10^{-5}$ \\
     \multicolumn{1}{c}{}  & (859$\times$) & (172$\times$) & (86$\times$) \\ \midrule
     \multicolumn{1}{c}{40 bits}  & {$4*10^{-6}$} & {$4.26*10^{-5}$} & {$ 8.8*10^{-5}$} \\
    \multicolumn{1}{c}{}  & {(214K$\times$)} & {(44K$\times$)} & {(22$\times$)} \\ \midrule
         \multicolumn{1}{c}{48 bits}  & {$8.2*10^{-11}$} & {$2*10^{-9}$} & {$8*10^{-9}$} \\
        \multicolumn{1}{c}{}  & ($152\times$) & ($755\times$) & ($1501\times$) \\
        \toprule
      \multicolumn{1}{c}{Random guess} \\
 \multicolumn{1}{c}{of $k \in D$} & {$5.4*10^{-13}$} & {$2.7*10^{-12}$} & {$5.4*10^{-12}$} \\ \bottomrule
\end{tabular}
\caption{\label{table:vulnerability_surf_real} Probability of a \emph{FindFPK} random guess finding an exploitable false-positive key against SuRF-Real. In parentheses, the
relative reduction (in \# of queries) of \attack compared to brute force search. The last row shows the probability of brute force guessing finding a full dataset key.
}
\end{center}
\end{table}

}

\section{Prefix Bloom filter \attack} \label{sec:attack-pbf}

This section instantiates \attack against LSM-trees using the prefix Bloom filter~(PBF)~\cite{PBF}.
We describe the PBF in~\cref{sec:pbf-primer} and show its vulnerability in~\cref{sec:pbf-vuln}.

\subsection{Prefix Bloom filter primer} \label{sec:pbf-primer}

The PBF is a Bloom filter-based range filter that supports range queries for ranges expressible as fixed-prefix queries.
While PBFs do not provide general range queries, they are currently deployed in real-world key-value stores such as RocksDB~\cite{RocksDB} and LittleTable~\cite{LittleTable}.

A PBF consists of a Bloom filter and a predetermined prefix length, $l$.
When a key $k$ is inserted into the PBF, both $k$ and its $l$-bit prefix are inserted into the Bloom filter.

PBF range queries must be for ranges of the form ``all keys starting with $\alpha$,'' where $\alpha$ is an $l$-bit string.
They are answered by querying the Bloom filter for $\alpha$.
If this query responds false, the dataset does not contain keys within the target range.

The PBF answers point queries by querying the Bloom filter for the queried key.
We remark that if the high-level system does not prioritize point query efficiency, the PBF can be configured to only store key prefixes.
In this case, the PBF implements a point query for key $k$ by querying its Bloom filter for $k$'s $l$-bit prefix.
This option reduces the PBF's memory consumption but increases the FPR of point queries.
This PBF configuration does not affect the success of our attack, so we do not discuss it further.

\subsection{Vulnerability of the PBF} \label{sec:pbf-vuln}

The PBF has the characteristics defined in~\cref{subsubsec:vulnrableFilters}.
As with SuRF, \Cref{char:distinguish}(1) holds trivially.
\Cref{char:fpk}(2) holds because the PBF's FPR is based on its Bloom filter's FPR.

The PBF has an important property: it not only has the usual Bloom filter false positives caused by hash collisions but also has what we call \emph{prefix false positives}.
These occur when a PBF point query falsely returns positive for an input $\kappa$ that is an $l$-bit prefix of a dataset key, simply because the Bloom filter stores both dataset keys and their $l$-bit prefixes.
This property implies that~\cref{char:1} holds: with probability $1-FPR$, an $l$-bit false-positive is actually the prefix of some stored key.

To show that~\cref{char:2} holds, we need only describe how to find prefix false positives~(\cref{sec:pbf-fpk}).
Finding them makes the \emph{IdPrefix} algorithm of~\cref{char:2} trivial: given an $l$-bit false positive $\kappa$, it outputs $\kappa$.

\subsubsection{Finding $l$-bit false-positive keys (FindFPK)} \label{sec:pbf-fpk}

The \emph{FindFPK} algorithm first determines the length of key prefixes stored in the PBF, $l$, and then proceeds to guess prefix false positives.
Crucially, finding $l$ needs to be performed only once per attack. That is, when running the attack's rounds concurrently~(\cref{sec:impl}), we run this step only once.

Once $l$ is known, generating queries for uniformly random $l$-bit strings will find false-positive keys, similarly to the SuRF attack's \emph{FindFPK}~(\cref{sec:surf-fpk}).
Given a set of false positive $l$-bit keys thus found, an expected fraction of $p/2^l$ will be prefix false positives, where $p$ is the number of distinct $l$-bit prefixes of dataset keys.
The remaining false positives will be hash-collision Bloom filter false positives.
Because we cannot distinguish between the two types of false positives, the attack's later steps must try to extend all of them to full keys.

The crux of the \emph{FindFPK} algorithm is to identify $l$.
To this end, we rely on the PBF property that made it vulnerable in the first place.
For any prefix length $l' \neq l$, the probability of an $l'$-bit key being a false positive is exactly the filter's FPR.
Only for $l$-bit keys will we observe a ``bump'' in the probability of a random $l$-bit key being a false positive, due to the presence of prefix false positives.

Accordingly, the \emph{FindFPK} algorithm first generates $j$ queries for uniformly random keys of length $l'$, for every non-trivial prefix length $l'$ (e.g., $l' \geq 3$).
It observes the fraction of false positives found and deduces that $l$ is the length $l$' for which the fraction of false positives found is maximal.

\FullPaper{
\subsection{Complexity analysis}

As with the SuRF's attack analysis, we consider a worst-case scenario of uniformly random keys.
Consider a dataset $D$ of $m$-bit uniformly random keys.
We compare the expected number of queries \attack requires to disclose a key to the number of queries required by brute force guessing.
Let $p \leq |D|$ be the number of distinct $l$-bit prefixes of dataset keys.
This means that \emph{FindFPK} finds a prefix false positive after $2^l/p$ queries on average (it does not try to avoid repetitions).
During these queries, however, it also finds $\varepsilon \cdot 2^l/p$ Bloom filter false positives on average.
Because it cannot distinguish a prefix false positive from a Bloom filter false positive, the attack goes on to invest an average of $(2^{m-l}-1)/2 \approx 2^{m-l-1}$ queries in trying to extend each such false positive into a key in step~\circled{3} of~\cref{sec:attack:step2}~(see~\cref{sec:stats_sc}).

Overall, the average cost of extracting a key in \attack is $C_p=(1+\varepsilon 2^l/p)\cdot 2^{m-l-1}$ queries.
In comparison, as explained in~\cref{sec:stats_sc}, brute force enumeration or guessing require $C_b=(2^m-|D|)/(|D|+1) \approx 2^{m}/|D|$ queries on average to guess a true positive key.

\Attack is more efficient than brute force when $C_p / C_b < 1$.
Because $p = \alpha |D|$ for some $0 < \alpha \leq 1$, we have $C_p / C_b = |D|/2^l + \varepsilon/\alpha$ which is $< 1$ when $|D|/2^l < 1-\varepsilon/\alpha$.
\Attack's profitability thus grows as $\varepsilon$ decreases and/or $\alpha$ increases.
This implies that \attack is very effective in the realistic scenarios of $|D| \ll 2^l$ (e.g., $|D|=500\,M$ and $l=40$), as in these cases, $\alpha \approx 1$.
}

\ShortPaper{
\section{Complexity analysis}

The key factor determining \attack's effectiveness is the probability of \emph{FindFPK} (step~\circled{1} in~\cref{sec:attack:step2}) guessing an \emph{exploitable} key $k$, which is a false positive whose longest shared prefix with stored keys is of length $l$, where $l$ is a predetermined constant for which extending $k$ into a full key is feasible~(step~\circled{3}).

The full version of this paper~\cite{full-version} includes a theoretical analysis of the SuRF and PBF attacks, which is omitted here due to space constraints.
We analyze the case of uniformly random keys, which is the worst case for our attack.
(If the key distribution is skewed, then~(1)~the guessing and full-key extraction steps can incorporate this knowledge; and~(2)~the prefixes SuRF stores are longer, so our attack will identify longer prefixes and thus extend them to full keys faster.)

The analysis derives the probability of \emph{FindFPK} guessing an exploitable key. This determines the expected number of queries to guess an exploitable key or, equivalently, the number of keys we ultimately expect to extract after investing $G$ guesses in \emph{FindFPK}.
These values also allow comparing the cost (in queries) of \attack to brute force search.

Under the realistic constraint that $|D| \ll 2^l$, where $D$ is the dataset (e.g., $|D|=500$\,M and $l=40$), we find that~(1)~\attack becomes more effective with growth in dataset size and better FPR---i.e., as the LSM-tree becomes more effective, so does \attack; and~(2)~\attack takes several orders of magnitude fewer queries to extract a key than an exhaustive brute force search.
}

\section{Implementation issues} \label{sec:impl}

In previous sections, we assume the attacker can check if a key is a filter positive or negative key, based on measuring query response times.
Here, we describe our implementation of this check.

The basic idea is simple. \Attack's preliminary phase~(\cref{sec:attack:step1}) derives a response time cutoff.
Keys whose query response time is below this cutoff are considered negative; otherwise, they are considered positive.
However, this cutoff only distinguishes between queries satisfied from memory and those involving I/Os.
Once a query for a false-positive key completes, the I/O it performs reads the relevant SSTable into the in-memory page cache.
Future queries for false-positive keys covered by this SSTable will thus get satisfied from memory.

To overcome this problem, we exploit the fact that the attack targets some production system, which is assumed to sustain heavy I/O load due its legitimate operation.
This property implies that if the attacker \emph{waits} after performing a false positive query, the SSTable brought in will be evicted from the page cache due to legitimate I/O traffic.

Unfortunately, waiting for even a few seconds after every query would make the attack impractical.
We solve this challenge by performing attack rounds in a \emph{concurrent, breadth-first} manner, as described below,
instead of working depth-first (finding a false-positive key and proceeding to identify its prefix and then to extract the full key).

Step~\circled{1} of~\cref{sec:attack:step2} (\emph{FindFPK} execution) generates $N$ random keys (false positive candidates) and measures a four-query average response time for each key to identify false-positive keys.
The averages are computed in a breadth-first manner: there are four iterations, each of which performs one query for each key.
Waiting for page cache evictions is done only between each iteration.

Step~\circled{2} (\emph{IdPrefix}) similarly executes iteratively, interleaving the next step of \emph{IdPrefix} for each false-positive key in each iteration, until all invocations output a prefix.
Again, waiting for page cache evictions is only done between iterations.

Step~\circled{3} (key extraction) likewise interleaves the searches extending each prefix.
We optimize step~\circled{3}'s general-case brute force suffix search by leveraging the fact that step~\circled{2} outputs a set of prefixes.
This enables us to discard short prefixes, so that step~\circled{3} only attempts to extend prefixes where the suffix search is feasible.

The interleaved execution of each step can be sped up using multi-core parallelization by assigning each core a subset of the $N$ random keys, false-positive keys,
or prefixes when executing step~\circled{1}, \circled{2}, and \circled{3}, respectively, in the above described manner.
This results in linear speedup (in the number of cores) of step execution time.
Our implementation parallelizes step~\circled{3}, whose execution time dominates the attack~(\cref{subsubsec: w_w.o_counters}), over 16 cores and leaves the other steps single-threaded.

\section{Evaluation}\label{sec:evaluation_sc}

In this section, we evaluate \attack attacks on SuRF and PBF in RocksDB.
We demonstrate the attack's feasibility, successfully mounting it against a full-fledged RocksDB key-value store employing SuRF~(\cref{subsec: Attack in Practice}).%
\footnote{We use the SuRF's authors' implementation, \url{https://github.com/efficient/SuRF}.}
We empirically analyze the SuRF attack's efficiency and sensitivity to data store size and filter FPR~(\cref{subsec: Attack Behavior}).
Consistent with our theoretical analysis, we find that the attack becomes more effective with growth in dataset size and better FPR---i.e., as the LSM-tree becomes more effective, so does \attack.
Finally, we demonstrate the attack against the PBF~(\cref{subsec:pbf-eval}).

\subsection{Experimental setup\label{subsec: system setup}}

Both clients and the attacked key-value store run on the same server.
However, the time differences we exploit can be measured over the network using prior techniques~(see~\cref{sec:threat}).

We use a server with two Intel Xeon Gold 6132 v6 (Skylake) processors, each of which has 14 2.6 GHz cores with two hyperthreads per core.
The server is equipped with 192\,GB DDR4 DRAM and two 0.5\,TB NVMe SSDs.
The server runs Ubuntu 18.04 and code is compiled with GCC 4.8.

\paragraph{RocksDB setup}
We use a version of RocksDB modified by the SuRF authors to employ SuRF~\cite{SuRF}.
The target RocksDB instance uses the NVMe devices as secondary storage.
We use Linux \texttt{cgroup} to limit RocksDB's available DRAM to 2\,GB.
This configuration emulates an industrial-scale, I/O heavy key-value store setup, in which storage capacity far exceeds DRAM capacity.

The RocksDB engine stores 64-bit keys and 1000-byte values and the SuRF-Real variant.
Unless noted otherwise, we use a datastore of 50\,M uniformly random keys (generated using SHA1).
We invoke RocksDB LSM-tree compaction after populating the datastore.
We do this to emulate the compaction that naturally occurs in a real workload due to insertions, because our experiments perform only \emph{get()}s.

\paragraph{Background load}
In all experiments, we emulate a realistic, loaded system by running 32 threads that constantly perform \emph{get()} queries for random keys, with 50\% of the queries targeting stored keys and 50\% targeting non-present keys.

\subsection{RocksDB+SuRF-Real key extraction} \label{subsec: Attack in Practice}

We implement the attack as described in~\cref{sec:impl}.
\Cref{subsubsec: Time DIff} evaluates the attack's first phase~(\cref{sec:attack:step1}), demonstrating that query response times can be used to distinguish negative from positive keys in practice, even in the presence of heavy background load.
\Cref{subsubsec: w_w.o_counters} evaluates the attack's second phase, which extracts full keys, and compares it to a brute force search.

\subsubsection{Negative/positive query time differences} \label{subsubsec: Time DIff}

In this phase, the attacker performs 10\,M \emph{get()} queries for randomly generated keys to build the response time distribution.
\Cref{tab:timing} shows the distribution of response times in terms of 5\,microsecond buckets.
The distribution is extremely skewed toward values $< 25$\,$\mu$s, which our attack therefore assumes are associated with negative keys.

\begin{figure}[t]%
  \centering
  \vspace{-12pt}
  \subfloat[{\footnotesize Buckets $\geq 25$\,$\mu$s: Absolute number of queried keys}]{\includegraphics[width=0.49\columnwidth]{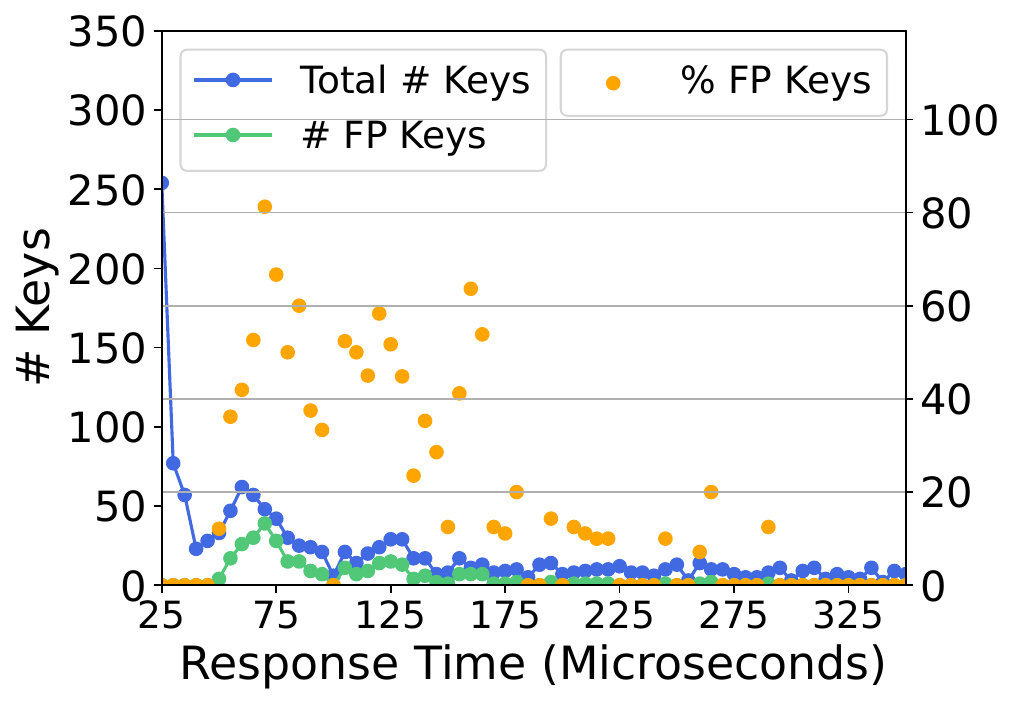}}\,
  \subfloat[{\footnotesize All buckets: Percentage of queried keys.}]{\includegraphics[width=0.49\columnwidth]{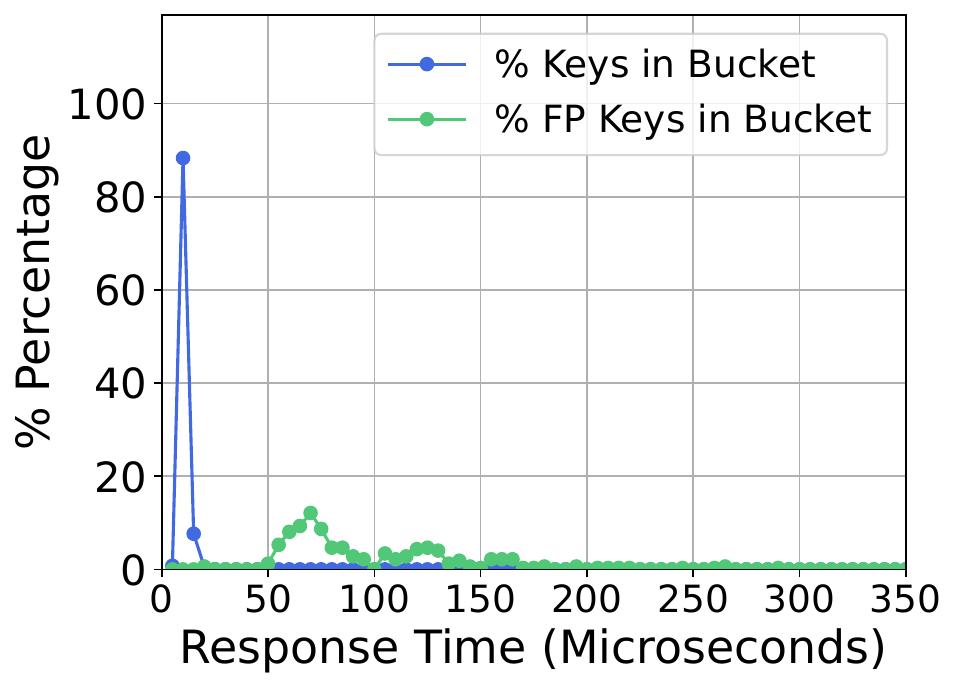}}%
  \vspace{-5pt}
  \caption{Breakdown of query response time distribution.}%
  \label{fig:timing}%
\end{figure}

To validate this assumption, \cref{fig:timing} visualizes the distribution while breaking the response times by queried key type (negative or false-positive).
This breakdown is presented for analysis purposes; it is not available to the attacker.
For readability, we present the breakdown in two ways.
\Cref{fig:timing}(a) shows only the buckets $\geq 25$\,$\mu$s, which are otherwise dwarfed by the lower end of the distribution.
We show both the number of keys ({\color{RoyalBlue}blue}) and false-positive ({\color{Emerald}green}) keys in each bucket, and the percent of false-positive keys in each bucket ({\color{YellowOrange}orange}).
\Cref{fig:timing}(b) shows the entire distribution, but bucket sizes (Y axis) are percentages instead of absolutes.
For each bucket, we report the percentage of keys in the bucket as well as the percentage of false positives (out of all positives).

\Cref{fig:timing}(a) shows that the vast majority of false positive queries have a response time of $>50$\,$\mu$s.
Conversely, \Cref{fig:timing}(b) shows that this response time range contains over 98\% of the false-positive keys.
Overall, these results show that picking a cutoff point of 50\,$\mu$s for distinguishing a negative from a positive key---which is done based only on the distribution's shape, without knowledge of key types---yields a good distinguisher.

\begin{table}[t]%
  \centering
  \footnotesize
    \begin{tabular}{lccl|}\toprule
         \textbf{Bucket range} & \textbf{\% of responses} \\
          \textbf{(microseconds)} &  \\ \toprule
         \multicolumn{1}{c}{$< 5$} & 0.77\% \\ \midrule
        \multicolumn{1}{c}{$\textbf{5 - 10}$} &  \textbf{88.3\%}\\ \midrule
        \multicolumn{1}{c}{$10 - 15$} & 7.65\%\\ \midrule
         \multicolumn{1}{c}{$15 - 20$} & 0.53\%\\ \midrule
        \multicolumn{1}{c}{$20 - 25$} & 0.05\%\\ \midrule
         \multicolumn{1}{c}{$\geq 25$} & 2.7\%\\ \bottomrule
    \end{tabular}
  \vspace{-5pt}
  \caption{Distribution of query response times.}%
  \label{tab:timing}%
\end{table}

\subsubsection{Key extraction} \label{subsubsec: w_w.o_counters}

The attack executes as described in~\cref{sec:impl}; specifically, wait is set to 20 seconds and each step is executed in a parallel, breadth-first manner, to minimize the amount of time spent waiting for page cache evictions.
The attacker generates a set of 10\,M random keys to find false-positive keys (step~\circled{1} of~\cref{sec:attack:step2}).
The attacker next identifies the prefix each false-positive key shares with a stored key (step~\circled{2}).
Finally, the attacker discards every prefix of length $< 40$\,bits
and attempts to extend the remaining prefixes into full keys~(step~\circled{3}).

\begin{figure}[t]
\centerline{\includegraphics[width=.8\columnwidth]{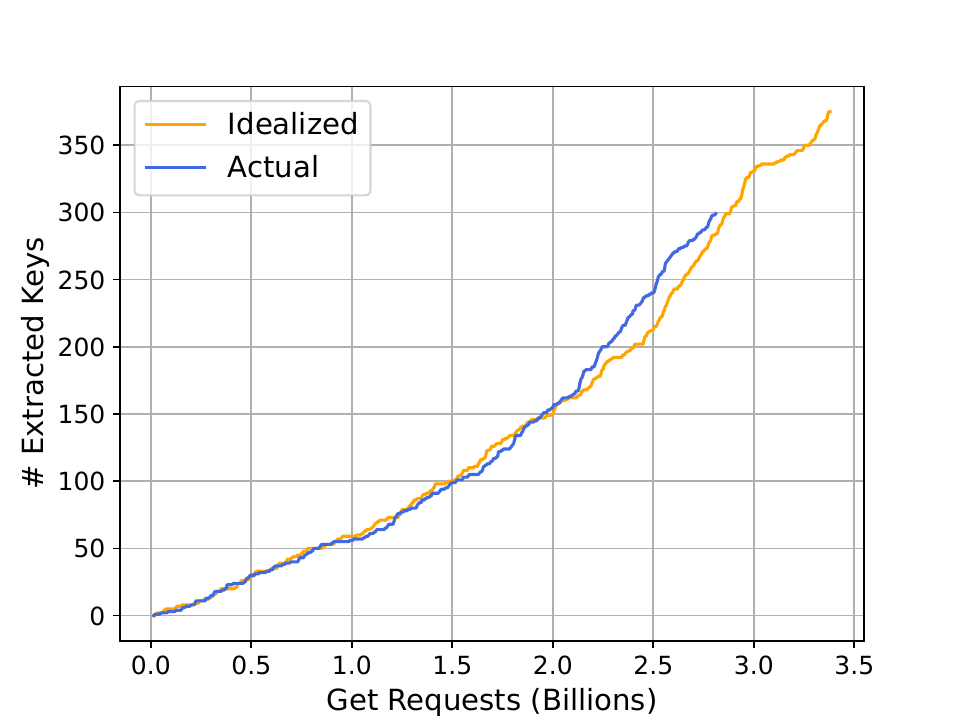}}
    \caption{Actual vs. idealized \attack against SuRF-Real: Number of keys extracted as attack progresses.}
   \label{fig:w_wo_counters1}
\end{figure}

\Cref{fig:w_wo_counters1} shows the number of keys extracted as a function of the number of total number of \emph{get()} requests issued by the attack (aggregated over steps~\circled{1}--\circled{3}).
The figure also compares the attack to an \emph{idealized} attack, which uses internal RocksDB debugging counters to accurately determine the filters' responses for each queried key, instead of relying on query response times.

Because the idealized attack never incorrectly classifies a key, it identifies more false positives than the actual attack in step~\circled{1}.
It thus requires more queries in step~\circled{2} to identify the shared prefixes of the keys provided to step~\circled{2}, as there are more of them.
Consequently, the idealized attack begins step~\circled{3} later (in terms of queries) than the actual attack, which is why its line is ``shifted'' compared to actual attack.
For this reason, the idealized attack also requires more queries overall.
Ultimately, however, the actual attack extracts only 74 fewer keys than the idealized version.

The idealized attack is also faster (in real time) than the actual attack, because it does not require waiting for page cache evictions.
The actual attack's key extraction rate is $\approx$~10~minutes/key, while the idealized attack achieves 0.2~minutes/key.

\Cref{table:gets_ditribution} shows a breakdown of the (actual) attack's queries across all three steps.
The bulk of the attack is spent on step~\circled{3}, extending prefixes into full keys.
Our later analysis~(\cref{sec:sens-size}) explains this number.
The table also reports wasted queries, which are issued when the attack futilely tries to extract a key from an incorrect prefix, which was misidentified due to incorrectly classifying a key as a false-positive~(based on its query response time).
Additional wasted queries~(not shown) are spent identifying prefixes of length~$<40$\, bits in steps~\circled{1}--\circled{2}, which are then discarded. While over 90\% of prefixes identified by steps~\circled{1}--\circled{2} are discarded, this waste is negligible, as they are discarded before the most expensive step.

\begin{table}[t]
\footnotesize
\begin{center}
\begin{tabular}{lccl}\toprule
      \multicolumn{1}{c}{\textbf{attack step}} &
      \multicolumn{1}{c}{\textbf{\# queries (millions)}} &
     \multicolumn{1}{c}{\textbf{queries/total (\%)}} \\
     \bottomrule
     \multicolumn{1}{l}{\circled{1} Find false positives} &	10M	& 0.35\% \\
     \midrule
     \multicolumn{1}{l}{\circled{2} Identify prefixes}	& 0.025M & 0.0009\% \\
      \midrule
     \multicolumn{1}{l}{\circled{3} Extract keys}  & 2581M & 91.68\%   \\
     \midrule
     \multicolumn{1}{l}{Wasted} & 224M & 7.9\%  \\ \bottomrule

\end{tabular}
\caption{Attack queries per stage. Wasted queries futilely attempt to extend an incorrectly identified prefix into a full key.}
\label{table:gets_ditribution}
\vspace{-25pt}
\end{center}
\end{table}

\paragraph{Comparison to brute force}
We further evaluate a brute force attack, that randomly guesses keys until a stored key is found.
We allow this attack to run for $10\times$ more time than the \attack experiment---but it fails to guess a key.
Unsurprisingly, brute force search for a large key space is infeasible.

\paragraph{SuRF-Hash vs. SuRF-Real}
SuRF-Hash complicates the attack.
Compared to SuRF-Real with the same per-key space budget, SuRF-Hash replaces key bits (SuRF-Real's suffix bits) with hash value bits.
This means that possible prefixes to identify are shorter and that the filter's FPR is lower, making the number of false positives identified in step~\circled{2} lower.
On the other hand, as discussed in~\cref{sec:surf-extend}, when identifying the prefixes and performing key extraction, the attacker can use the false-positive key's hash value to ignore definitely incorrect guess---potentially improving the attack's efficiency.

\begin{figure}[t]
\centerline{\includegraphics[width=.8\columnwidth]{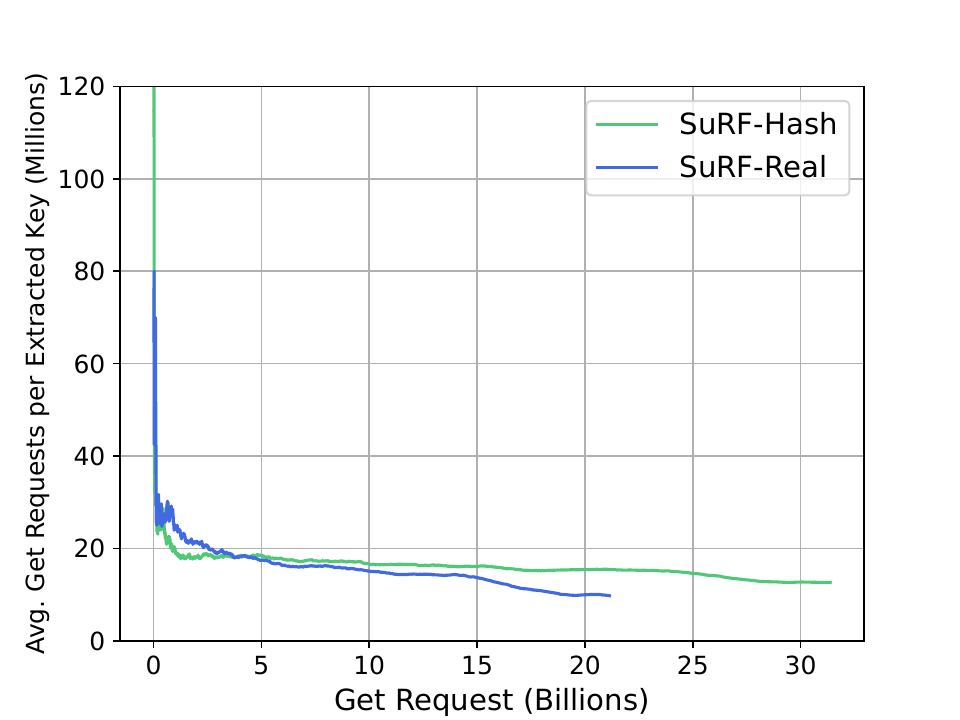}}
    \caption{SuRF-Hash vs. SuRF-Real: Moving average of queries per extracted key as a function of attack progress (measured in queries).}
\vspace{-5mm}
    \label{fig:HashgetsPerKey}
\end{figure}

To evaluate this trade-off, we compare idealized attacks against the same dataset, with RocksDB using either SuRF-Real with 8-bit suffixes or SuRF-Hash with 8-bit hashes.
Thus, in SuRF-Hash, the suffix search space when extracting a key $256\times$ larger than in SuRF-Real, but the attacker will ignore $255/256$ of its guesses on average.
To compensate for SuRF-Hash's lower FPR, the initial false-positive key search of the SuRF-Hash attack uses $3\times$ the number of candidate keys used for SuRF-Real.
\Cref{fig:HashgetsPerKey} therefore compares the attacks' amortized cost, in terms of a moving average of queries per extracted key as a function of attack progress.%
\footnote{I.e., the Y axis reports the number of \emph{get()}s issued divided by the number of keys extracted up to the current X-axis point.}
The SuRF-Hash attack's extra initial queries (for finding false positives) manifest as the peak of the per-key cost, when all these extra queries are amortized across only a handful of keys.
The extra cost is eventually amortized away, into a per-key cost of ~12\,M vs. 10\,M queries for SuRF-Hash vs. SuRF-Real, respectively.
For this similar cost, the SuRF-Hash attack extracts 2490 keys vs. 2171 keys for the SuRF-Real attack.

\subsection{Attack analysis} \label{subsec: Attack Behavior}

This section analyzes the attack's efficiency~(\cref{subsubsec: Efficiency}) and sensitivity to data store size~(\cref{sec:sens-size}) and filter FPR~(\cref{sec:sens-fpr}).

\subsubsection{Efficiency} \label{subsubsec: Efficiency}

\begin{figure}[t]
\centerline{\includegraphics[width=.8\columnwidth]{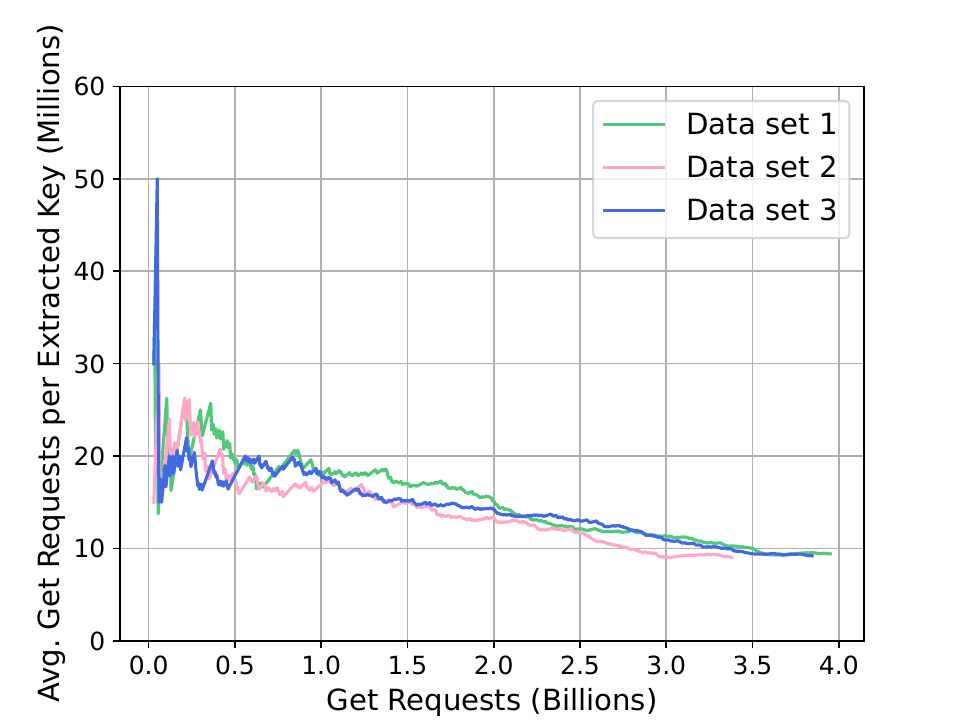}}
    \caption{Attack efficiency: average number of \emph{get()}s per extracted key as attack progresses.}
    \label{fig:getsPerKey}
\end{figure}

\Cref{fig:getsPerKey} shows the attack's efficiency, measured as average \emph{get()}s per extracted key as a function of attack progress.
We compare across three 50\,M random 64-bit key sets to show the results are not a function of the specific key set.

The average number of queries per extracted key converges to about 9\,M $\approx 2^{23}$.
This indicates that the attack extracts keys with roughly the work required to search a 23-bit space---$40992\times$ better than a brute force search of the full key space ($2^{64}/50\textrm{\,M}\approx 2^{38.4}$).
The attack also extracts a substantial number of keys (375, 419, and 423 keys).

\subsubsection{Sensitivity to dataset size} \label{sec:sens-size}

To evaluate the attack's sensitivity to the dataset size, we progressively shrink our original 50\,M key set into smaller subsets of size $c\cdot 10$\,M keys for $c \in [1,5]$.
We then perform an idealized attack against the system with each dataset, but using the same set of random keys for step~\circled{1}, so any difference in attack behavior can be related only to the datastore size and not the key distribution.

\begin{figure}[t]
\centerline{\includegraphics[width=.8\columnwidth]{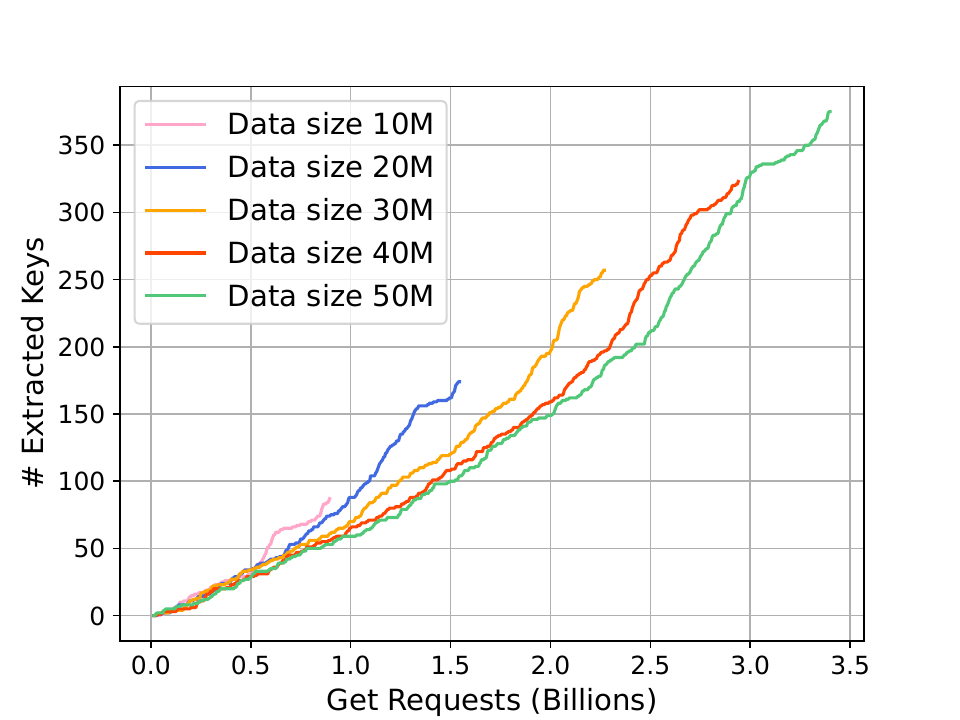}}
    \caption{Idealized attack against SuRF-Real: Number of keys extracted for different dataset sizes.}
    \label{fig:filterSize}
\end{figure}

\Cref{fig:filterSize} shows the number of keys extracted as the attack progresses.
\Attack is more effective as the dataset size increases: it extracts $\approx 100$ keys from the 10\,M dataset, but almost 400 keys from the 50\,M dataset.

\subsubsection{Sensitivity to SuRF FPR} \label{sec:sens-fpr}

We show that \attack becomes more effective as SuRF's FPR improves, i.e., the attack becomes more harmful to the system as SuRF becomes more productive to the system.
To demonstrate this effect, we compare idealized attacks against the same dataset, with RocksDB using either SuRF-Base or SuRF-Real.
SuRF-Base stores shared key prefixes, padded to the next full byte (which adds 1--8 bits to the prefix).
SuRF-Real does the same, plus stores a byte from the key's unique suffix, and thereby improves its FPR~(see~\cref{sec:surf-primer}).

\begin{figure}[t]
\centerline{\includegraphics[width=.8\columnwidth]{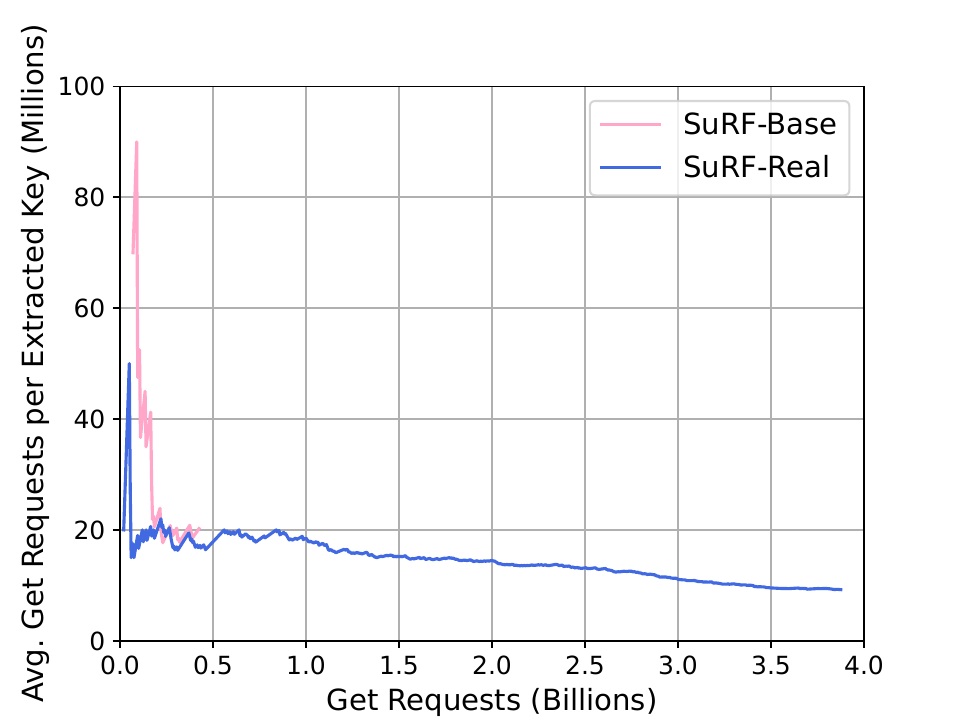}}
    \caption{SuRF-Real vs. SuRF-Base: Moving average of queries per extracted key as a function of attack progress (measured in queries).}
    \label{fig:BaseRealgetsPerKey}
\end{figure}

We carry out the attacks against each SuRF variant using the same initial random key set, used to identify false-positive keys.
\Cref{fig:BaseRealgetsPerKey} reports the attack's amortized cost (queries per extracted key) as the attack progresses.

In both cases, the attack has similar efficiency of $\approx 10$\,M queries per extracted key, as evident from the similar slope of the two lines.
However, the attack is more successful against SuRF-Real, where it extracts 420 keys, than against SuRF-Base, where it extracts 21 keys.
The reason for the improved effectiveness is that SuRF-Real's extra key byte storage makes an initial false-positive key much more likely to have a prefix length of $>40$\,bits,
resulting in more false positives making it to step~\circled{3}.

The situation is similar with SuRF-Hash, which further improves the FPR over SuRF-Real~(\cref{fig:HashgetsPerKey}).
As mentioned in~\cref{subsubsec: w_w.o_counters}, the idealized SuRF-Hash attack extracts 2490 keys vs. 2171 keys for the idealized SuRF-Real attack.

\subsection{RocksDB+PBF key extraction} \label{subsec:pbf-eval}

\begin{figure}[t]
\centerline{\includegraphics[width=.8\columnwidth]{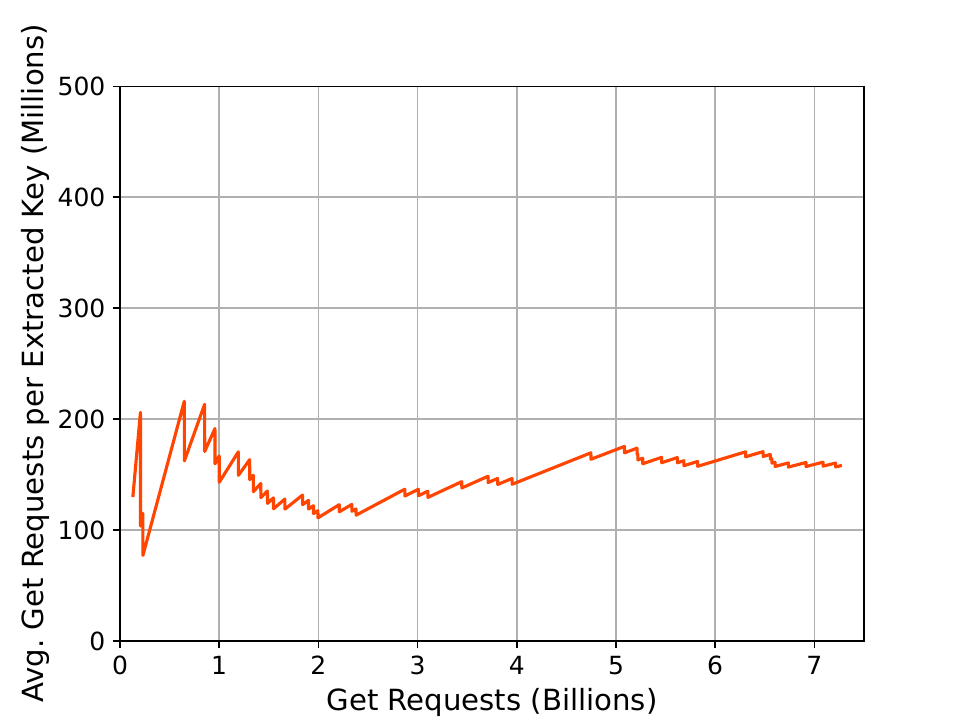}}
    \caption{Idealized \attack against PBF ($l=40$ bits).}
   \label{fig:prefix_bloom_attack}
\end{figure}

We evaluate an idealized \attack attack against RocksDB's PBF.
We use a dataset of 50\,M uniformly random 64-bit keys.
We configure the PBF to store prefixes of length $l=40$ bits and to consume 18 bits/key (which is roughly the space usage of SuRF in our experiments).

Step~\circled{1} (\emph{FindFPK}) perform 1\,M queries for uniformly random 40-bit keys, which result in 457 false-positive keys.
The attack then attempts to extend these false positives into full keys. It eventually extracts 46 keys, which matches the expected number of prefix false positives observed in 1\,M random guesses ($1\textrm{M} \cdot 50\,\textrm{M}/2^{40} = 45.4$).
\Cref{fig:prefix_bloom_attack} plots the attack’s amortized cost (queries per extracted key) as the attack progresses.
The PBF attack makes 160\,M queries per extracted key, which is $20\times$ more queries/key than the SuRF attack, but still three orders of magnitude better than a brute force search.
The reason for this difference is that the PBF attack wastes effort trying to extend Bloom filter false positives that are not prefix false positives.

\section{Mitigation} \label{sec:mitigation}

Here, we discuss approaches for mitigating \attack attacks.
Unfortunately, every potential solution constitutes some trade-off, whether in query performance, memory efficiency, complexity, or other system aspects.

\paragraph{System-level approaches}
A system can block \attack attacks by only querying its key-value storage engine for keys the requesting user is allowed to access.
This approach requires re-architecting the system so that a key's ACL is kept outside of the key-value store.
In addition, a system can rate limit user requests, thereby slowing down \attack attacks. This approach is viable only if the system is not meant to handle a high rate of normal, benign requests.

\paragraph{Key-value store mitigation}
A key-value engine can block \attack by maintaining separate filters for point and range queries for each SSTable file. Unfortunately, this approach will double filter memory consumption. In addition, it will not block attacks that target range queries (which we believe are possible, and are currently exploring).

\paragraph{Filter-level mitigation}
A natural mitigation is for key-value stores to employ non-vulnerable range filters.
Like the separate filter approach described above, this mitigation carries the risk of being vulnerable to future extensions of \attack to range queries.

In addition, the properties that make a range filter non-vulnerable to point query-based \attack may limit its utility in practice.
For example, Rosetta (Robust Space-Time Optimized Range Filter)~\cite{Rosetta} is a range filter that does not conform to our vulnerable range filter characterization~(\cref{subsubsec:vulnrableFilters}), but it lacks support for variable-length keys, which are important in practice.

Rosetta uses Bloom filters for SuRF-like prefix-based filtering.
Rosetta assumes a bound on the possible key length in bits, $L$.
A Rosetta instance consists of $L$ Bloom filters, $B_1,\dots,B_L$.
When a key $k$ is inserted into the filter, each $i$-bit prefix is inserted into the $i$-th Bloom filter $B_i$.
A Rosetta point query thus simply queries $B_L$, making Rosetta non-vulnerable to \attack.

The Rosetta paper does not specify how variable-length keys are handled.
Its design is clearly incompatible with such keys if there is no predetermined bound on their size.
Even if such a bound exists (and can thus be used for $L$), Rosetta requires every key to be padded to $L$ bits, so that point queries function correctly. This requirement significantly increases the filter's memory consumption.

\paragraph{Encrypted key-value stores}
Disclosed keys reveal no sensitive information if they are stored encrypted in the storage engine.
However, encrypting key-value pairs requires re-architecting the entire system so it can query on encrypted data~\cite{EncKV,EncKV0}.
Most if not all deployed key-value stores do not support such encryption.

\section{Related Work}\label{sec:literature_sc}

\paragraph{Key-value store timing attacks}
Existing key-value store timing attacks aim to disclose stored values.
These attacks work by exploiting external mechanisms such as memory deduplication~\cite{KVDataAttack1} or memory compression~\cite{KVDataAttack2}, which can be disabled for protection.
In contrast, \attack exploits a mechanism of the key-value store itself, which cannot be disabled for protection without suffering significant throughput degradation and additional I/O traffic.

\paragraph{Storage engine timing attacks}
Timing attacks mostly target cryptographic software rather than storage engines.
Futoransky et al.~\cite{SideChannelBtree} extract private keys from a MySQL database with a timing attack, but the attack relies on insertions of attacker-chosen data.
Wang et al.~\cite{Side-ChannelSearchIndexes} show a practical timing attack on a multi-user search system, such as Elasticsearch.

\FullPaper{
\paragraph{Filter attacks}
Privacy attacks against Bloom filters have been explored in the context of privacy-preserving record linkage~\cite{kuzu2011constraint, kuzu2013practical,niedermeyer2014cryptanalysis, kroll2015, christen2017efficient,christen2018pattern, vidanage2019efficient, mitchell2017graph},
where Bloom filter serve as encodings of sensitive information.
These attacks reveal stored keys by exploiting skewness of the stored key distribution.
In contrast, \attack does not require skewness and succeeds on uniformly random keys.
}

\section{Conclusion}\label{sec:conclusion}

This paper shows that certain range filters make LSM-trees vulnerable to novel \emph{\attack} timing attacks, which exploit differences in query response times to reveal keys and prefixes of keys stored in the LSM-tree.
\FullPaper{Unlike prior key-value store timing attacks, \attack abuses internal mechanisms of the key-value store---range filters---which cannot be disabled for protection without suffering significant throughput degradation and additional I/O traffic.}
Our results show that key-value store performance improvements may trade security in exchange, and encourage practitioners and researchers to evaluate the security impact of their work.
We also hope that our characterization of vulnerable range filters will spur research on more secure filters.

\section*{Acknowledgments}
We extend our deepest thanks to Yuvraj Patel,
\ShortPaper{the paper's shepherd, and the anonymous reviewers }%
\FullPaper{the shepherd of the USENIX ATC'23 version of this paper, and that version's anonymous reviewers}
for their dedication and assistance in improving this paper and their valuable feedback.
We thank Guy Khazma for his work on an earlier stage of this project.

\bibliographystyle{plain}
\bibliography{main}

\end{document}